\documentclass[aps,prb,twocolumn,superscriptaddress]{revtex4}
\usepackage{graphicx}
\usepackage{amssymb}
\usepackage{amsmath}
\usepackage{amsfonts}

\begin{document}

\title{Towards an effective theory of skyrmion crystals} 
\author{V.~E. Timofeev}
\affiliation{NRC ``Kurchatov Institute", Petersburg Nuclear Physics Institute, Gatchina
188300, Russia}
\affiliation{St.Petersburg State University, 7/9 Universitetskaya nab., 199034
St.~Petersburg, Russia} 
\affiliation{St. Petersburg Electrotechnical University ``LETI'', 197376 St. Petersburg, Russia}
\author{A.~O. Sorokin}
\affiliation{NRC ``Kurchatov Institute", Petersburg Nuclear Physics Institute, Gatchina
188300, Russia} 
\author{D.~N. Aristov}
\affiliation{NRC ``Kurchatov Institute", Petersburg Nuclear Physics Institute, Gatchina
188300, Russia}
\affiliation{St.Petersburg State University, 7/9 Universitetskaya nab., 199034
St.~Petersburg, Russia} 
\affiliation{Institute for Nanotechnology, Karlsruhe Institute of Technology, 76021
Karlsruhe, Germany }
 
\begin{abstract}
We consider multiskyrmion configurations in 2D ferromagnets with Dzyaloshinskii-Moriya (DM) interaction and the magnetic field, using the stereographic projection method.  In the absence of DM interaction, $D$, and the field, $B$,  the skyrmions do not interact and the exact multiskyrmion  solution is a sum of individual projections.  In certain range of $B, D\neq0$, skyrmions become stable and form a hexagonal lattice.  The shape of one skyrmion on the plane is fully determined by $D$ and $B$. We describe  multiskyrmion configurations by simple sums of individual skyrmion projections, of the same shape and adjusted scale. This procedure reveals pairwise and triple interactions between skyrmions,  and the energy of proposed hexagonal structure is found in a good agreement with previous studies.   It allows an effective theory of skyrmion structures in terms of variables, referring to individual skyrmions, i.e., their position, size and phase, elliptic distortions etc.
\end{abstract}

\maketitle

There is a renewed interest to topological aspects of matter in recent years. In the context of magnetic substances, one of the object under intensive theoretical and experimental investigations is the topologically nontrivial spin texture called the skyrmion lattice. Such a texture may occur in two-dimensional magnets under various conditions (see \cite{Nagaosa13,Garst17} for a review), in particular, in the presence of an external magnetic field and either the Dzyaloshinskii-Moriya interaction (DMI)  \cite{Bogdanov89,Bogdanov94,Rossler06,Binz06}, or a geometrical frustration \cite{Kawamura12}, or four-spin exchange interactions \cite{Heinze11}. In the first case, the texture has a characteristic size more than the lattice constant and has been observed by different experimental methods \cite{Heinze11,Muhlbauer09,Yu10,Pfleiderer11}.

A mathematical description of such textures faces some difficulties. One difficulty in the description in terms of single skyrmions is associated with the stabilization of them. Strictly speaking, a single skyrmion remains unstable in the bulk even in the presence of the DMI and an external magnetic field, $B$. In particular, stabilization of a skyrmion requires that a spin configuration is homogeneous at infinity, but DMI leads to the modulated ground state \cite{Bak80,Nakanishi80}. Thus, skyrmions can only be considered as stable topologically non-trivial configurations either on a disk of finite radius or as an element of a lattice \cite{Bogdanov94}.

A multi-skyrmion configuration can be easily described in the absence of DMI and $B$ by the Belavin-Polyakov (BP) solution \cite{Belavin1975}, wherein skyrmions do not interact. However,  DMI  and the field not only change the shape of skyrmions, but also lead to the interaction between them. These interactions change the symmetry of the system and, as a consequence, the order parameter space from $G/H=SO(3)/SO(2)$ of the BP case to $G/H=SO(2)\otimes SO(2)$ \cite{Sorokin14}. This implies that a skyrmion lattice is better described as Abrikosov vortices rather than BP skyrmions (see also \cite{Babaev14,Garaud14,Schmalian15}).

Another possible way to describe a skyrmion lattice is based on an analogy with the one-dimensional helical magnet with the easy-plane anisotropy \cite{Aristov12,Borisov09,Kiselev12,Togawa12}, where a skyrmion lattice and a helix (deformed by an external field) are equivalent. The two-dimensional skyrmion lattice partly resembles (but is not equivalent to) a superposition of three helices directed at an angle of 120 degrees relative to each other \cite{Muhlbauer09}. The solution for this three-helix configuration, taking into account the above deformations, is yet to be found. Thus, despite difficulties discussed above, the representation of a skyrmion lattice in terms of single skyrmions turns to be the most useful from the theoretical point of view for considering this structure and its low-energy dynamics by analytical methods. As we observe below, the usefulness of this representation is supported by the fact that the shape of a skyrmion core remains nearly intact upon variation in a wide range of model parameters as well as upon small deformations of a skyrmion lattice.

In this paper we propose to describe a skyrmion configuration in terms of simple sum of stereographic projections of individual skyrmions, similar to \cite{Belavin1975}.
We first find the optimal size of skyrmions on a disc of finite radius with homogeneous boundary conditions. This solution is natural for a single skyrmion but inconvenient for the lattice. However, the relative intactness of the skyrmion core shape upon small deformations allows one to use the single skyrmion solution in construction of many-skyrmion configuration. The energy of the proposed trial function is in a good agreement with previous calculations.

We also discuss the interaction between two and three skyrmions, it turns out that the triple interaction cannot be ignored for realistic setups. The proposed construction allows conceptually simple analysis of the effective dynamics of the skyrmion lattice, with the collective variables describing various deformations of stereographic projections of individual skyrmions. The appearing expressions and peculiarities of the dynamics should be presented in a separate study.

\textbf{2.} We consider a continuum version of 2D ferromagnet characterized by the uniform exchange $C$,
Dzyaloshinskii-Moriya (DM) interaction, $D$,  and external magnetic field $B$. The classical energy of the model is $E = \int d\mathbf{r}\, \mathcal{H} $ with the energy density
\begin{equation}
\mathcal{H} =     \tfrac{1}{2} C \partial_{\mu}\varphi^{i}\partial_{\mu}\varphi^{i} - D
\epsilon_{\mu ij} \varphi^{i}\partial_{\mu}\varphi^{j}  + B(1 - \varphi^{3})  \,,
\end{equation}
where   $\varphi^{i}   = \langle S^{i}\rangle /S $ is  average local magnetization, $i=1,2,3$ and $\mu=1,2$,  $\epsilon_{\mu ij}$ is totally antisymmetric tensor; we set $C=1$ below.

At zero temperature the magnetization is normalized to unity, $\varphi^{i}\varphi^{i}=1$, so that $\vec{\varphi} = (\sin\theta \cos\phi, \sin\theta \sin\phi,\cos \theta)$. We choose the  stereographic projection approach and write
 \begin{equation}
\begin{aligned}
\varphi^{1} + i \varphi^{2} & = \frac{2f(z,\bar{z})}{1 + f(z,\bar{z})\bar{f}(z,\bar{z})}   \,, \\
\varphi^{3} &= \frac{1 - f(z,\bar{z})\bar{f}(z,\bar{z})}{1 + f(z,\bar{z})\bar{f}(z,\bar{z})}  \,, \\
\end{aligned}
\label{def}
\end{equation}
with  $f(z,\bar{z})$  a complex-valued  function of $z=x+iy$ and $\bar{z} = x - iy$.

\begin{figure}[t]
\center{\includegraphics[width=0.95\linewidth]{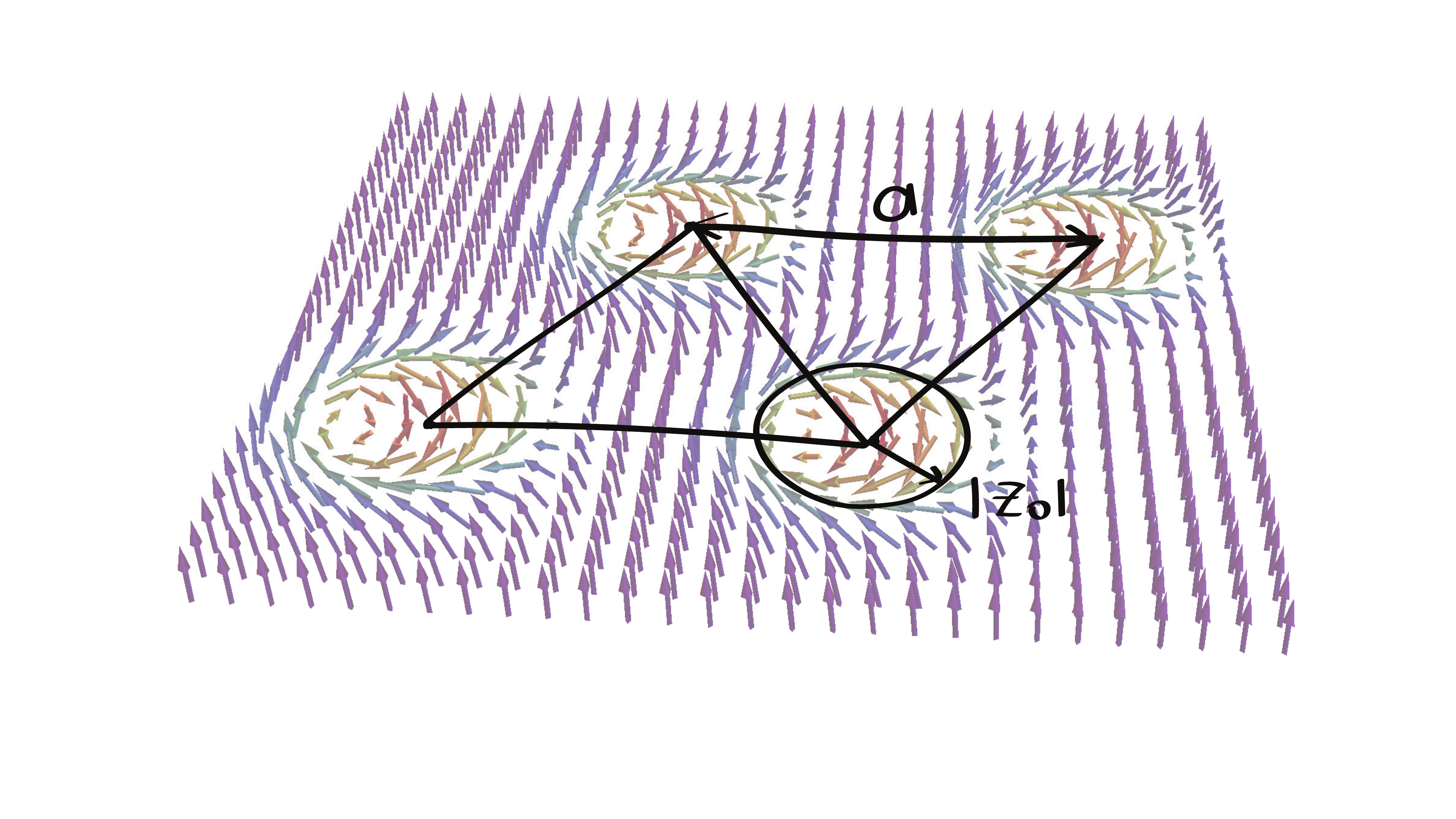}}
\caption{Skyrmionic crystal with hexagonal unit cell. Two  scales are shown: the period, $a$, and skyrmion  scale, $z_{0}$.  }
\label{fig:cell}
\end{figure}

Introducing  $\partial_{z} \rightarrow \frac{1}{2} (\partial_{x} - i \partial{y})$ and  $\partial_{\bar{z}} \rightarrow \frac{1}{2} (\partial_{x} + i \partial{y})$, the energy is written in new terms as  $E = \tfrac 12 \int d \mathbf{r} \, \mathcal{H} $ with
 \begin{eqnarray}
 \mathcal{H} [f]& =& \frac{4(\partial_{z}f\partial_{\bar{z}}\bar{f}+\partial_{z}\bar{f}\partial_{\bar{z}}f) }{(1 + f\bar{f})^{2}}  \label{Hviaf} \\ &
+& \frac{2iD (\bar{f}^{2}\partial_{\bar{z}}f + \partial_{\bar{z}}\bar{f}-\partial_{z}f - f^{2}\partial_{z}\bar{f})}{(1 + f\bar{f})^{2}}  + \frac{2Bf \bar{f}}{1 + f\bar{f}}    \nonumber \,,
\end{eqnarray}
and topological charge is given in terms of $f$ by
\begin{equation}
Q=\frac{1}{4\pi}\int{d \mathbf{r} \ \frac{4(\partial_{z}\bar{f}\partial_{\bar{z}}f -\partial_{z}f\partial_{\bar{z}}\bar{f})}{(1+f \bar{f})^{2}}}\,.
\label{topcharge}
\end{equation}
The Euler-Lagrange equations take the form
 \begin{equation}
\begin{aligned}
&    2f\partial_{z}\bar{f}\partial_{\bar{z}}\bar{f}
- (1 + f \bar{f})\partial_{z}\partial_{\bar{z}}\bar{f}
\\ &  -  iD(\bar{f}\partial_{\bar{z}}\bar{f} + f\partial_{z}\bar{f}) + \tfrac14 B\bar{f} (1 + f\bar{f}) = 0
\end{aligned}
\label{}
\end{equation}
This equation is strongly nonlinear and cannot be analytically solved in general.  However, in the simple case $D=B=0$ this nonlinear equation allows any holomorphic (or anti-holomorphic) function as a solution \cite{Belavin1975}.  Particularly, a single skyrmion solution corresponds to $f=z_{0}/\bar{z}$, with arbitrary complex $z_{0}$.
Multiskyrmion solution is given by a sum
\begin{equation}
  f(z) = \sum_{j} \frac{z_{0}^{(j)}}{\bar{z} - \bar{z_{j}} } \,.
  \label{BPsolution}
\end{equation}

Motivated by this observation we take a following trial form for one skyrmion at $D,B \neq 0$ :
\begin{equation}
f(z,\bar{z}) = \frac{e^{i \alpha} \kappa (z\bar{z})}{\bar{z}}  \,,
\label{Ansatz}
\end{equation}
with real valued function $\kappa (z\bar{z})$.
In BP case for single skyrmion $\kappa$ is constant and equals both to skyrmion scale and skyrmion size, as defined below.

In our case  $D,B\neq 0$ the function $\kappa$ is determined numerically and shows a smooth behavior and an exponential decay at large distances. The phase of skyrmion is found as $\alpha=  \frac{\pi}{2}\,  \mbox{sign } D$, with the sign corresponding to the sign of DM interaction;  it means that the magnetization direction is perpendicular to the vector $\mathbf{r}$ in plane.

Let us discuss now several spatial scales arising in our problem.
\\  i)
The principal spatial scale, $L=C/D$, is related to the pitch of the helix, which is formed in the film at low temperatures at zero external field.
\\ ii)
Another scale, which we call properly skyrmion's scale,  is the residue of the stereographic projection at the center of the skyrmion. This quantity depends on the distance between skyrmions and is denoted by $z_{0}$. 
\\  iii)
The size of the skyrmion may be defined by the condition $\langle \varphi^{3}\rangle =0$, or $|f(z,\bar{z})|=1$,  when the center of skyrmion is given by $\langle \varphi^{3} \rangle = -1$, i.e. the maximum projection $|f(z,\bar{z})|\to\infty$.
\\ iv)
The exponential decrease of the skyrmion tail is defined by yet another distance, the correlation length in uniform ferromagnet, $\ell = (C/B)^{1/2}$.
\\ v)
Finally, we have the period of the hexagonal lattice of skyrmion crystal, $a$, see Fig.\ \ref{fig:cell}

It turns out however that all these spatial scales are of the same order in the considered range of parameters of the model. In this sense we deal with skyrmions of small radius, in contrast to \cite{kravchuk2018spin}.


\textbf{3.}
The metastable BP solution \eqref{BPsolution} means that the skyrmions do not interact with each other at $D=B = 0$. In our case it is not so and the formation of static multiskyrmion configuration assumes several steps.
An individual skyrmion in the discussed range of parameters provides a gain in energy for some  $\kappa(x)$, this gain  $\mu \sim  C$ being the chemical potential. The multiskyrmion solution is eventually formed, with the distance between skyrmions, $a$, defined by the repulsion between them. This repulsion can decrease at the expense of certain increase of the energy of individual skyrmions with the modified function $\kappa$.
Ultimately we anticipate the formation of hexagonal skyrmion lattice, where each skyrmion is surrounded by six neighbors exerting pressure on it. This high-symmetry surrounding can be modeled by putting the skyrmion at the center of the disc, in order to determine the corresponding change in its shape function.

The equation for the profile function in the bulk is 
\begin{equation}
\kappa(B \kappa^{2} - 4\kappa (D + 2 \kappa') + x(B + 8  \kappa'^{2}))= 4x(x+\kappa^{2})\kappa''  \,,
\label{eq:kappa}
\end{equation}
with $x=z\bar{z}$, $\kappa'(x)=d\kappa(x)/dx$ etc.  
 In BP case, $B=D=0$, we obtain either skyrmion solutions, $\kappa \sim 1$, $\kappa \sim x$, or meron one,  $\kappa = x^{1/2}$.

We solve \eqref{eq:kappa} on a disc of large radii $R \gg L$ with the boundary condition $\kappa(R^{2})=0$ which is done by "shooting" method for moderate $R$.
The limiting form of $\kappa(x)$ at $R\to \infty$  is obtained by matching $\kappa(x)$ at smaller $x$ with the   large $x$ asymptotic.

Comparing the solutions obtained for different $R$ we found an interesting feature of $\kappa(x)$.
As $R$ decreases from $R=\infty$ to $R=4/D$, the function  $\kappa(x)$ varies in magnitude but not in its shape at moderate $x$, as we explain now. 

\begin{figure}[t]
 \includegraphics[width=0.95\columnwidth]{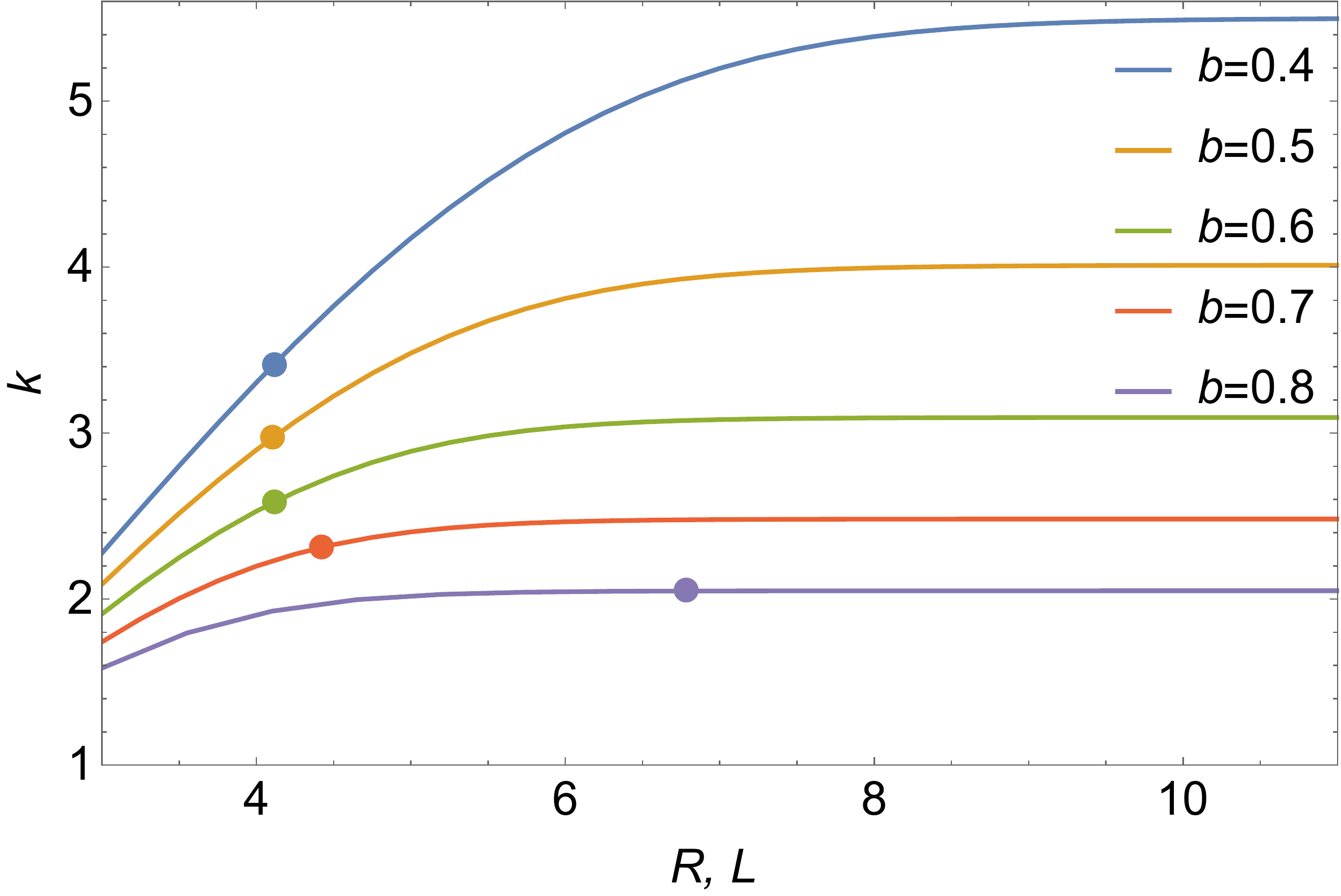}
\caption{The dimensionless residue $k$ for various field strength, $b$, is shown  as a function of the radius of the disc, $R$,  where the  skyrmion  is confined. Circles  on the curves correspond to disc radii providing the minimum of the energy density. }
\label{fig:KvsR}
\end{figure}

\begin{figure}[t]
 \includegraphics[width=0.95\linewidth]{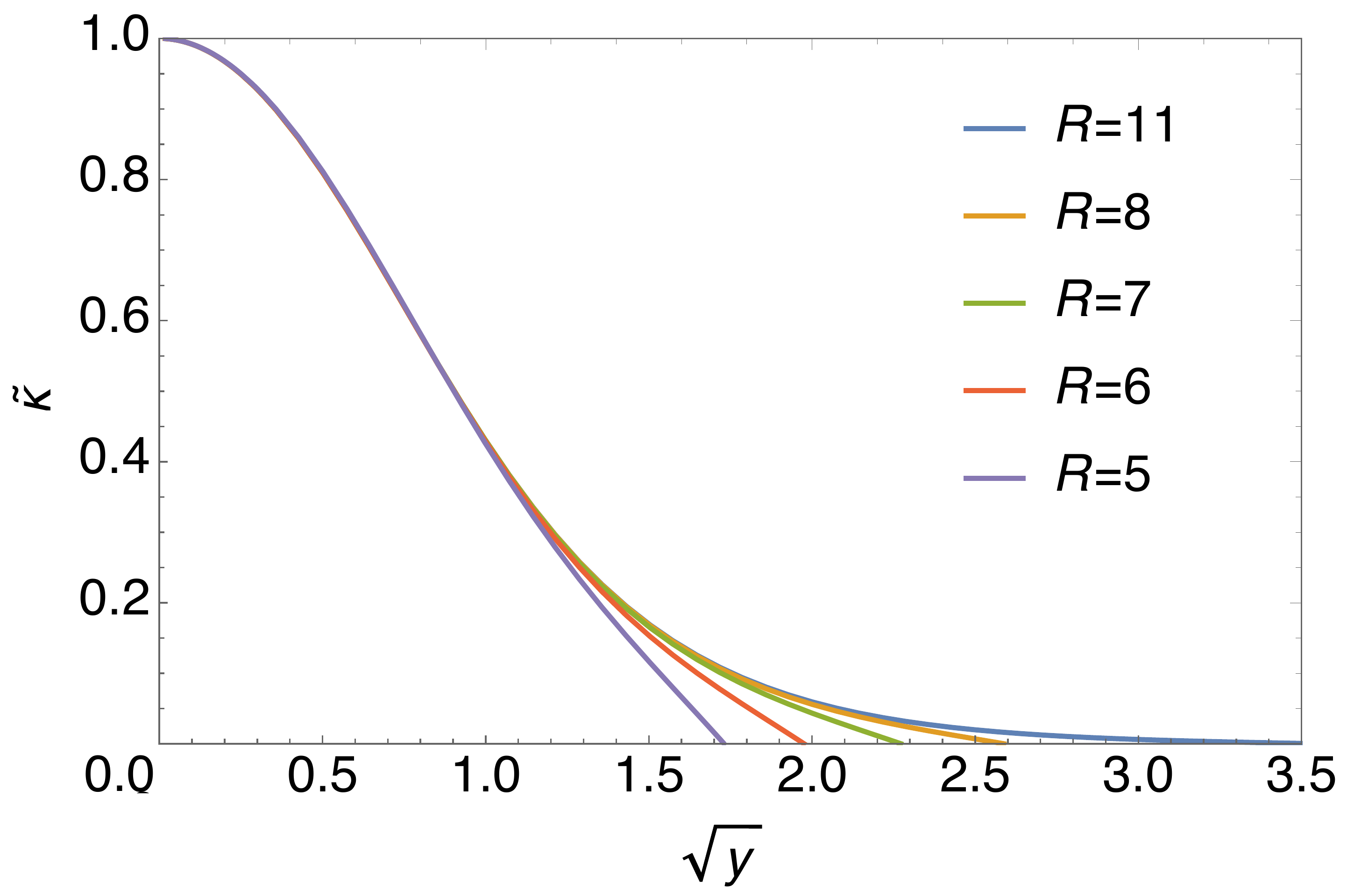}
\caption{\label{fig:scaling}The shape of rescaled modulating function $\widetilde{\kappa}(y)$,Eq.\ \eqref{rescale}, obtained for different disc radii $R$ and
$b=0.6$.  Nearly invariant form of $\widetilde{\kappa}(y)$ at smaller distances is visible.}
\end{figure}

We introduce 
the dimensionless field $b= BC/D^{2} = L^{2}/\ell^{2}$ and write the residue at $\bar z=0$ in Eq.\  \eqref{Ansatz}  in dimensionless form, $\kappa(0) = L k$.
Notice that $k$ is a function of disc radius, $R$, as is shown in Fig.\ \ref{fig:KvsR}.
Next we define a dimensionless function
\begin{equation}
\widetilde{\kappa}(y)= (kL)^{-1} \kappa\left(y\, k^{2} L^{2}  \right)
\label{rescale}
\end{equation}
which is shown in  Fig.\  \ref{fig:scaling} for different values of $R$.

%
%
It is  seen in this figure, that only a ``tail'' of solutions $\widetilde{\kappa}(y)$ for various disc radii is different, whereas their ``core'' is nearly unchanged.
This peculiar property of preserving the shape of the core of the skyrmion can be attributed to the  form of the equation \eqref{eq:kappa}. 
 Writing $\widetilde{\kappa}(y)= 1 + c_{1} y + c_{2} y^{2}/2+\ldots$
we find from  \eqref{eq:kappa}
 \begin{equation}
\begin{aligned}
 c_{1}  &=\tfrac18 k(bk -4 )  =
-\frac1{2b} +  \frac b8 \left(k-\frac2b \right)^{2} ,
 \\
 c_{2} &= \frac1{3b} +
 \frac{1}{48}\left(k-\frac2b \right)  \left(b^2 k^3-4 bk (k-1) +8\right)
\end{aligned}
\label{}
\end{equation}
It turns out  that in the considered range of $b\in (0.3,0.8)$ the values of optimal $k$, delivering the minimum of energy density, roughly satisfy the relation $bk \approx 2$, as can be checked in Fig.\ \ref{fig:KvsR}. It results in relative smallness of the second terms
in above coefficients $c_{1,2}$ so that  $c_{1} \simeq -\frac1{2b}$ , $c_{2} \simeq \frac1{3b} $, which leads to quasi-invariant scaling form of $\widetilde{\kappa}(y)$ at $y\lesssim 1$ shown in Fig. \ref{fig:scaling}.
To a good accuracy we can approximate $\widetilde{\kappa}(y) \simeq \exp(-y/(2b))$, i.e. Gaussian shape since $y/b \sim |z|^{2} \ell^{2}/L^{4}$.
At larger distances, $y\gg1$, this quasi-invariance  is lost, as the true  asymptote reads $\widetilde{\kappa}(y) \sim  y^{1/4}  \exp(-k\sqrt{by})$;  this asymptotic behavior is unimportant for us below.
The skyrmion size, $r$, defined by the condition $|F\left ( {r}/{z_{0}} \right )| =1$, is always smaller than $|z_{0}|$  since $\tilde \kappa(x) \leq 1$ ; we find that $r$ varies from  $r  \simeq 0.6 |z_{0}|$ at $b=0.2$  to $r\simeq 0.72 |z_{0}|$ at $b = 0.8$.

We use the quasi-invariance property of $\widetilde{\kappa}(y)$ in the mostly important region of $ r \lesssim L $ below.
First we determine  the function $\kappa(x)$ for given $b$ in the limit of large disc radius, $R/L \to \infty$, and the corresponding limiting function $\tilde \kappa_{\infty}(y)$.
Then we define our ansatz function for the individual skyrmion as (cf.\ \eqref{Ansatz})
\begin{equation}
f(z , \bar z) =  F\left (\frac{\bar z}{z_{0}} \right ) \equiv
\frac{z_{0}}{\bar z} \tilde \kappa_{\infty}\left ( \left|\frac{\bar z}{z_{0}}\right |^{2} \right)  \,,
 \label{eq:skyform1}
\end{equation}
with   the complex-valued scale,  $z_{0}$, considered as a variational parameter.
Eqs.\ \eqref{Hviaf}, \eqref{eq:skyform1}
lead to the quadratic form for the energy
\begin{equation}
E_{1} (z_{0})= C (a_{1}  - a_{2} \mbox{Im }(z_{0})/L +   a_{3} |z_{0}|^{2}/\ell^{2} ) \,,
\label{minZ0}
\end{equation}
whose minimum, by construction, is provided by purely imaginary $z_{0}$
with $|z_{0}| = k _{R=\infty}L $ the scale  of a single skyrmion on the plane.


\textbf{4.}
Based on the quasi-invariance of the shape of a single skyrmion 
we propose to model the multi-skyrmion configuration by the sum
\begin{equation}
 f (z)= \sum_{j}F\left ( ({\bar z  -\bar z_{j}})/{z_{0}^{(j)}} \right )  \equiv  \sum_{j}{f_{j}} ,
 \label{ansatz}
\end{equation}
with $z_{j}$ the center of $j$th skyrmion and $z_{0}^{(j)}$ its scale.
In case $D=B=0$ we recover the formula \eqref{BPsolution}.
We consider below the case of skyrmions of the same scale $z_{0}^{(j)}= z_{0}$, purely imaginary quantity.

The form \eqref{Ansatz} automatically gives the topological charge $Q=1$ for one skyrmion 
The sum \eqref{ansatz} of stereographic projections  of $N$ skyrmions corresponds to topological charge $N$, which is easy to show for $\tilde \kappa\equiv 1$ and can be proved in general by performing continuous deformation of $\kappa(z\bar z)$ upon which the charge is not changed. This property is also verified numerically.

The interaction between skyrmions is given by a difference between the energy of the whole configuration and sum of energy of the parts:
\begin{equation}
U=\mathcal{H} \left [ \sum_{j}{f_{j}} \right] - \sum_{j} \mathcal{H}[f_{j}]
\end{equation}
The expression \eqref{Hviaf} for  $\mathcal{H}[f]$ is quite complicated and besides pairwise interaction results also in sizable triple interaction of three skyrmions.  The latter interaction may be important for stabilization of the  skyrmion lattice.

\begin{figure}[t]
\center{\includegraphics[width=0.95\linewidth]{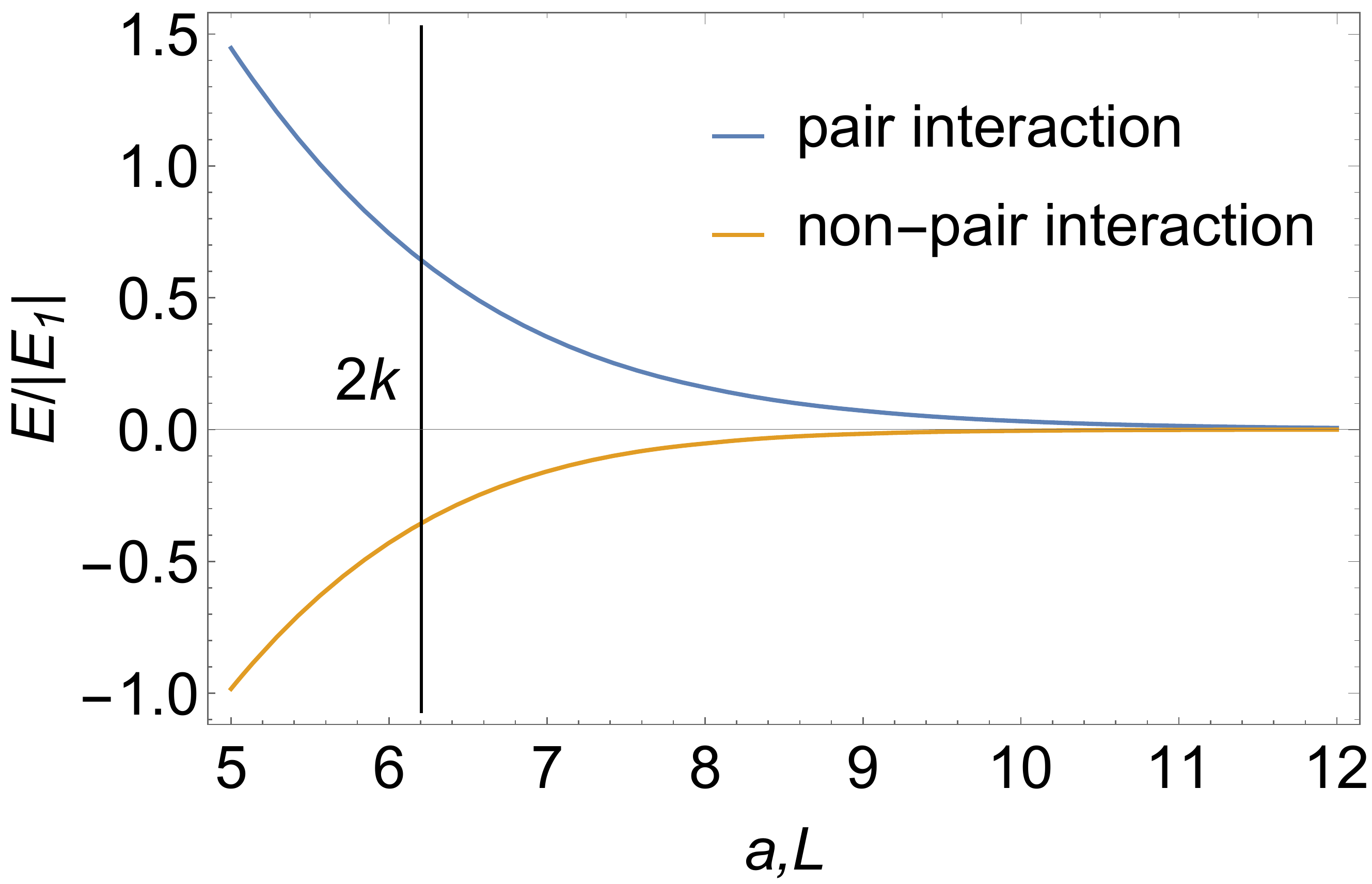}}
\caption{Pair and non-pair interaction of skyrmions with $z_{0}=k$ for $b=0.6$. Vertical line show values of interaction with cell parameter equal two radius of free single skyrmion"$2k$".
}
\label{fig:interaction}
\end{figure}

Let us  first discuss the pair interaction. Placing two skyrmions of scale $z_{0}$ at the distance $a$ and fixing their positions, we determine the value of the interaction
\begin{equation}
U_{2}(z_{0},a) =\mathcal{H} \left [  f_{1} + f_{2}  \right] - \mathcal{H}[f_{1} ]  - \mathcal{H}[f_{2} ]
\end{equation}
If the radii of the skyrmions are taken equal to their radii in solitary situation, then the interaction between them is repulsive and quite large, $\sim C$,  for $R\sim |z_{0}|$, see Fig.\ \ref{fig:interaction}.
For fixed $a$ it is therefore energetically favorable to reduce the size of individual skyrmions, thus enhancing their energy, but lowering their interaction.

We find that the energy of two skyrmions  is approximately given  in units of $C$ by (cf. \eqref{minZ0})
\begin{eqnarray}
E_{2}&
\simeq & 2 \tilde a_{1} + 2\tilde a_{3} (x-1)^{2} + u_{2} x^{4} \exp{(2r_{0}-a) /\ell} \,,
\label{energy2}
\\
 \tilde a_{1}  &=&   a_{1} -  \tilde a_{3} \,,\quad \tilde a_{3} = \frac{a_{2}^{2}}{4ba_{3}}  \,,
 \quad
 x = |z_{0}|/r_{0} \,,  
 \nonumber
\end{eqnarray}
with $\tilde a_{1}\simeq -5$, $\tilde a_{3} \simeq 20 $, $u_{2} \simeq 3.6$ for typical $b=0.6$; the last term in \eqref{energy2} corresponds to $U_{2}(z_{0},a)$.

The triple interaction can be calculated similarly, it depends on several distances between skyrmions. To simplify the discussion, we determine the value of non-pairwise interaction by placing skyrmions at vertices of equilateral triangle. We skip the results for three skyrmions, and   focus on the four-skyrmion configuration.  Motivated by other studies, we consider a primitive cell of a triangular lattice depicted in Fig.\ \ref{fig:cell}, with lattice period $a$ and the the radii of skyrmions, $|z_{0}|$, corresponding to their optimal value in solitary configuration.
Non-pairwise interaction  can then be found in the form
\begin{equation}
U_{3} (a) =\mathcal{H} \left [ \sum\limits_{j=1,\ldots4}{f_{j}} \right] -4 \mathcal{H}[f_{1}] - 5 U_{2}(a)
\end{equation}
which turns out to be sizable and attractive as shown in Fig.\ \ref{fig:interaction}.
We find that the energy of four skyrmion configuration in Fig.\ \ref{fig:cell} is given to a good accuracy by the energy of two adjacent triangles, and the interaction at the distances $\sim \sqrt{3} a$ between the furthest skyrmions in  Fig.\ \ref{fig:cell}  is negligible.

It allows us to extend our treatment to the periodic structure of triangular lattice.  We should take into account that each skyrmion in Fig.\ \ref{fig:cell}  belongs to four cells, and four out of five links also belong to adjacent cells.  It leads to the definition of the energy per unit cell
 \begin{equation}
\begin{aligned}
 E_{cell} =  &    \mathcal{H}[f_{1}] +3 U_{a}(a)+  U_{3}(a)   \\
\end{aligned}
\end{equation}


Calculating the energy density per unit cell and comparing it to the density obtained for the disc of optimal radius we see a very good agreement, which validates the choice of our ansatz \eqref{ansatz}.

\textbf{5.}
The obtained formulas can be used for description of the low-energy action of a skyrmion lattice.
One possibility to do it is to change the argument in one of the functions,  $F\left (\frac{\bar z}{z_{0}} \right ) \to F\left (\frac{\bar z + \delta \bar z}{z_{0}} \right )$ with
\begin{equation}
\delta \bar z = \varepsilon_{0} \bar z + \sum _{m\geq1} (\varepsilon_{m} \bar z^{m+1} + \varepsilon_{-m}  z^{m-1})
\end{equation}
and complex-valued $\varepsilon_{m}$. Both dilatation of the skyrmion and the change of its phase are encoded in $ \varepsilon_{0}$,   the shift of its position is $ \varepsilon_{-1}$,   the special conformal transformation in BP case \cite{aristov2015magnon} is given by  $ \varepsilon_{1}$.  Elliptic distortions  \cite{Bogdanov94a} correspond to $ \varepsilon_{\pm2}$.

The dynamics \cite{Sheka2001} can be analyzed by inclusion of the kinetic part of the Lagrangian, $\sim \dot \phi \cos\theta$. One finds the effective action by variation in $\epsilon_{m}$ and considering only the quadratic terms \cite{Metlov}.
It shows that the phase of  $\varepsilon_{m}$ is roughly a canonically conjugated momentum to its absolute value,  $|\varepsilon_{m}|^{2}$. This can be most easily shown for the  $\varepsilon_{0}$ mode, by allowing the dilatations of only one skyrmion. In general the situation is much more involved, because even at the level of deformations of one skyrmion, the collective variables $\varepsilon_{m}$ and $\varepsilon_{-m}$ are coupled. 
Consideration of deformations of several skyrmions, $\delta \bar z_{j}$, should lead to the effective theory  in terms of collective variables $\varepsilon_{m}^{(j)}$ on the hexagonal lattice.  Such analysis is rather involved and is delegated to future publication.
\medskip

We thank B.A. Ivanov, K. Metlov, M. Garst, and A.~Tsypilnikov for useful discussions and communications.


\begin{thebibliography}{27}

\bibitem{Nagaosa13} N. Nagaosa and Y. Tokura, Nature Nanotech. {\bf8}, 899 (2013).

\bibitem{Garst17} M. Garst, J. Waizner, and D. Grundler, J. Phys. D: Appl. Phys. {\bf50}, 293002 (2017).

\bibitem{Bogdanov89} N. Bogdanov and D.A. Yablonskii, Sov. Phys. JETP {\bf68}, 101 (1989).

\bibitem{Bogdanov94} N. Bogdanov and A. Hubert, J. Magn. Magn. Mater. {\bf138}, 255 (1994).



\bibitem{Rossler06} U.K. Ro\ss ler, N. Bogdanov, C. Pfleiderer, Nature {\bf442}, 797 (2006).

\bibitem{Binz06} B. Binz, A. Vishwanath, and V. Aji, Phys. Rev. Lett. {\bf96}, 207202 (2006).

\bibitem{Kawamura12} T. Okubo, S. Chung, and H. Kawamura, Phys. Rev. Lett. {\bf108}, 017206 (2012).

\bibitem{Heinze11} S. Heinze et al., Nature Phys. {\bf7}, 713 (2011).

\bibitem{Muhlbauer09} S. M\"uhlbauer et al., Science {\bf323}, 915 (2009).

\bibitem{Yu10} X.Z. Yu et al., Nature {\bf465}, 901 (2010).

\bibitem{Pfleiderer11} C. Pfleiderer, Nature Phys. {\bf7}, 673 (2011).

\bibitem{Bak80} P. Bak and M.H. Jensen, J. Phys. C: Solid State Phys. {\bf13}, L881 (1980).

\bibitem{Nakanishi80} O. Nakanishi, A. Yanase, A. Hasegawa, and M. Kataoka, Solid State Commun. {\bf35}, 995 (1980).

\expandafter\ifx\csname natexlab\endcsname\relax\def\natexlab#1{#1}\fi
\expandafter\ifx\csname bibnamefont\endcsname\relax
  \def\bibnamefont#1{#1}\fi
\expandafter\ifx\csname bibfnamefont\endcsname\relax
  \def\bibfnamefont#1{#1}\fi
\expandafter\ifx\csname citenamefont\endcsname\relax
  \def\citenamefont#1{#1}\fi
\expandafter\ifx\csname url\endcsname\relax
  \def\url#1{\texttt{#1}}\fi
\expandafter\ifx\csname urlprefix\endcsname\relax\def\urlprefix{URL }\fi
\providecommand{\bibinfo}[2]{#2}
\providecommand{\eprint}[2][]{\url{#2}}

\bibitem{Belavin1975}
\bibinfo{author}{\bibfnamefont{A.~A.} \bibnamefont{Belavin}} \bibnamefont{and}
  \bibinfo{author}{\bibfnamefont{A.~M.} \bibnamefont{Polyakov}},
  \bibinfo{journal}{JETP Lett.} \textbf{\bibinfo{volume}{22}},
  \bibinfo{pages}{245} (\bibinfo{year}{1975}).

\bibitem{Sorokin14} A.O. Sorokin, JETP {\bf118}, 417 (2014).

\bibitem{Babaev14} J. Garaud, K.A.H. Sellin, J. J\"aykk\"a, and E. Babaev, Phys. Rev. B {\bf89}, 104508 (2014).

\bibitem{Garaud14} D.F. Agterberg, E. Babaev, and J. Garaud, Phys. Rev. B {\bf90}, 064509 (2014).

\bibitem{Schmalian15} M.S. Scheurer and J. Schmalian, Nature Commun. {\bf6}, 6005 (2015).

\bibitem{Aristov12} D.N. Aristov and A. Luther, Phys. Rev. B {\bf65}, 165412 (2002).

\bibitem{Borisov09} A.B. Borisov, J. Kishine, I.G. Bostrem, and A.S. Ovchinnikov, Phys. Rev. B {\bf79}, 134436 (2009).

\bibitem{Kiselev12} V.V. Kiselev and A.A. Raskovalov, Theor. Math. Phys. {\bf173}, 1565 (2012).

\bibitem{Togawa12} Y. Togawa et al., Phys. Rev. Lett. {\bf108}, 107202 (2012).

\bibitem{kravchuk2018spin} V.P. Kravchuk et al.  Phys. Rev. B {\bf97}, 064403 (2018).

\bibitem{aristov2015magnon}  D.N. Aristov, S.S. Kravchenko, and A.O. Sorokin, JETP lett. {\bf102}, 455 (2015).

\bibitem{Bogdanov94a} N. Bogdanov and A. Hubert, phys. stat. sol. (b) {\bf186}, 527 (1994).


\bibitem{Sheka2001}  D. D. Sheka,  B. A. Ivanov, and F. G.  Mertens,
Phys. Rev. B {\bf 64}, 024432 (2001)


\bibitem{Metlov} K.L. Metlov, Phys. Rev. B {\bf88}, 014427 (2013).


\end{thebibliography}
\end{document}